# *P*-shell carriers assisted dynamic nuclear spin polarization in single quantum dots at zero external magnetic field

C. F. Fong,[1,*]   Y. Ota,[2]   E. Harbord,[1,3]   S. Iwamoto,[1,2]   and Y. Arakawa[1,2,†]

[1] *Institute of Industrial Science, The University of Tokyo, 4-6-1 Komaba, Meguro-ku, Tokyo 153-8505, Japan*

[2] *Institute for Nano Quantum Information Electronics, The University of Tokyo, 4-6-1 Komaba, Meguro-ku, Tokyo 153-8505, Japan*

[3] *Centre for Nanoscience and Quantum Information, University of Bristol, Tyndall Ave, Bristol, Avon BS8 1FD, United Kingdom*

Repeated injection of spin polarized carriers in a quantum dot leads to the polarization of nuclear spins, a process known as dynamic nuclear spin polarization (DNP). Here, we report the first observation of *p*-shell carrier assisted DNP in single QDs at zero external magnetic field. The nuclear field - measured by using the Overhauser shift of the singly charged exciton state of the QDs - continues to increase, even after the carrier population in the *s*-shell saturates. This is also accompanied by an abrupt increase in nuclear spin buildup time as *p*-shell emission overtakes that of the *s*-shell. We attribute the observations to *p*-shell electrons strongly altering the nuclear spin dynamics in the QD, supported by numerical simulation results based on a rate equation model of coupling between electron and nuclear spin system. DNP with *p*-shell carriers could open up avenues for further control to increase the degree of nuclear spin polarization in QDs.

[*] cffong@iis.u-tokyo.ac.jp
[†] arakawa@iis.u-tokyo.ac.jp

# I. INTRODUCTION

.

Semiconductor quantum dots (QD) confine carriers in all three spatial dimensions, giving rise to strongly coupled electron-nuclear spin systems in which interactions are mediated by the hyperfine interaction.[1,2] As a result, electron spins can be transferred to the nuclear spins via a mutual spin "flip-flop" process. Continuous injection of spin-polarized electrons polarizes the nuclear spin ensemble - generating a nuclear field - in a process known as dynamic nuclear spin polarization (DNP). The feedback of DNP has led to the observation of surprising effects such as the enhanced degree of spin polarization in charged excitons,[3–5] and bistability of the nuclear field with respect to excitation power,[6–9] polarization of optical excitation[6] and external magnetic field.[10,11] Also consequences of the back-action of DNP such as line dragging effects where the QD resonance is "locked" to the laser excitation have been observed,[12,13] as well as the narrowing of nuclear spin fluctuation with two-laser excitation.[14,15]

Prior to previous reports of DNP at zero external applied magnetic field,[4,16] it was generally assumed that a nonzero external magnetic field was necessary to produce polarized nuclear spins. Lai *et al.* proposed that DNP at zero external field was possible, as the effective inhomogeneous magnetic (Knight) field generated by optically excited electrons is larger than the local nuclear field fluctuations, pre-empting the need for an external field. Dzhioev and Korenev suggested that the nuclear quadrupole interaction is more likely to be responsible for DNP at zero external field as the depolarization of the nuclei via the dipole-dipole interaction is supressed.[17]

In previous experiments, non-resonant or quasi-resonant excitation creates carriers which rapidly relax to the ground state energy levels (*s*-shell) of the QD, where these carriers interact with the nuclear spins[3–11,16–18] prior to radiative or non-radiative recombination. While the contribution of the first excited state (*p*-shell) electrons to DNP has been suggested in previous work,[19] it has not been studied so far. Here we demonstrate the first *p*-shell electron assisted DNP at zero external magnetic field. We observed a continued increase in the nuclear field even after the saturation of the *s*-shell states, as well as an abrupt increase in the nuclear spin buildup time, $T_{\text{buildup}}$, after the closing of the *s*-shell. These results can be interpreted in terms of *p*-shell electron orbitals, in which high spatial variation of *p*-shell electron wave functions can support a strong inhomogeneous Knight field, slowing the nuclear spin decay. These interpretations are supported by simulations which investigate the effects of nuclear spin polarization rate and decay rate on the overall nuclear field.

## II. SAMPLE PREPARATION AND EXPERIMENTAL SETUP

The sample under investigation is grown by molecular beam epitaxy on a (001) GaAs substrate. A single InAs QD layer is capped with an 80-nm-thick GaAs layer. Atomic force microscopy analysis of uncapped samples gave an estimated QD areal density of $\sim 5 \times 10^8$ cm$^{-2}$. This sample is subjected to rapid thermal annealing. The details of the growth conditions can be found elsewhere.[20,21] The sample is patterned with 1 µm diameter mesas by e-beam lithography followed by dry etching, in order to perform single QD spectroscopy with the following micro-photoluminescence setup.

A continuous wave (CW) semiconductor laser operated at 785 nm is focused on the sample with an objective lens (50×, NA = 0.65). The sample is held in a cryostat at a temperature of 7 K. The laser excites carriers non-resonantly above the GaAs bandgap and due to optical selection rules, a maximum carrier degree of polarization of 50% can be introduced into the QDs,[2] allowing us to generate spin majority carriers. The emitted PL is subsequently collected by the same objective lens and is analyzed with a computer controlled rotating quarter wave plate (QWP), followed by a linear polarizer, before being dispersed with a spectrometer and detected with a CCD. The linear polarizer is fixed and the QWP rotated, in order to avoid effects arising from the anisotropic polarization response of the spectrometer.

To measure the nuclear spin buildup time, an electro-optic modulator is driven by an appropriate square wave electrical signal to alternate the polarization of the excitation laser between right ($\sigma+$) and left ($\sigma-$) circular polarization over a range of 10 Hz – 50 kHz. This generates electrons in the GaAs that are majority polarized spin up and spin down respectively. The electron spins are transferred to the nuclei such that they generate nuclear fields of alternating polarities. At each frequency, a resultant nuclear field is generated and is reflected in the relative shift of the emission peak energy known as the Overhauser shift (OS). We used the emission from charged exciton states in the *s*-shell, $X^{+/-}$, as probes of the nuclear spin polarization in our QDs since the excitons couple to the light field and exhibit an OS even in the absence of any external magnetic field.[4,18,22] The emission of the QD is collected over an integration time of 1-3s in order to ensure that the QD is excited by a sufficient number of cycles of the polarization modulation to achieve dynamic equilibrium.

**III. OPTICAL CHARACTERIZATION AND NUCLEAR SPIN BUILDUP TIME**

Figure 1 shows the PL spectrum of the single QD under investigation, with peaks corresponding to *s*-shell and *p*-shell carrier recombination at a high excitation power of 2.0 μW as labelled. Here, DC excitation is used where the laser is set to a fixed polarization without modulation. The *p*-shell emission is identified by looking at the PL power dependence which was observed to have the characteristic super-linear increase.[23] The energy separation of the *p*-shell from the *s*-shell emission is about 40 - 50 meV which corresponds to the separation in the energy levels in a QD, consistent with previously reported values.[24]

We identify each excitonic complex in the *s*-shell by a combination of power and polarization dependent spectroscopy. Neutral excitons, $X^0$ and biexcitons, $XX^0$ show linear and quadratic power dependence respectively.[25,26] In addition, they exhibit equal and opposite fine structure splitting which arises due to the anisotropic electron-hole exchange interaction.[27] Charged excitons, on the other hand, have no fine structure splitting.[27] To distinguish between positive and negative charged states, the QD is pumped with CW fixed circularly polarized light, without polarization modulation. $X^{+(-)}$ couple to two orthogonal circularly polarized photons depending on the spin of the single photoexcited (resident) electron, as such giving light of different circular polarization after recombination. $X^+$ exhibits dominant co-polarized emission,[28] while $X^-$ shows dominant counter polarized emission,[29] allowing them to be unambiguously identified (Fig. 1(inset)). For the case of fixed polarization excitation with no modulation, the observed splitting or the OS, is the difference in the emission peak energy at the two orthogonal circular polarizations detection. The OS arises mainly from the *s*-shell electron–nuclear spin interaction since the hole is *p*-like with weak hyperfine interaction.[30] The nuclear field shifts the spin up(down) electron state to lower(higher) energy and recombination with the holes give photons of lower(higher) energy.

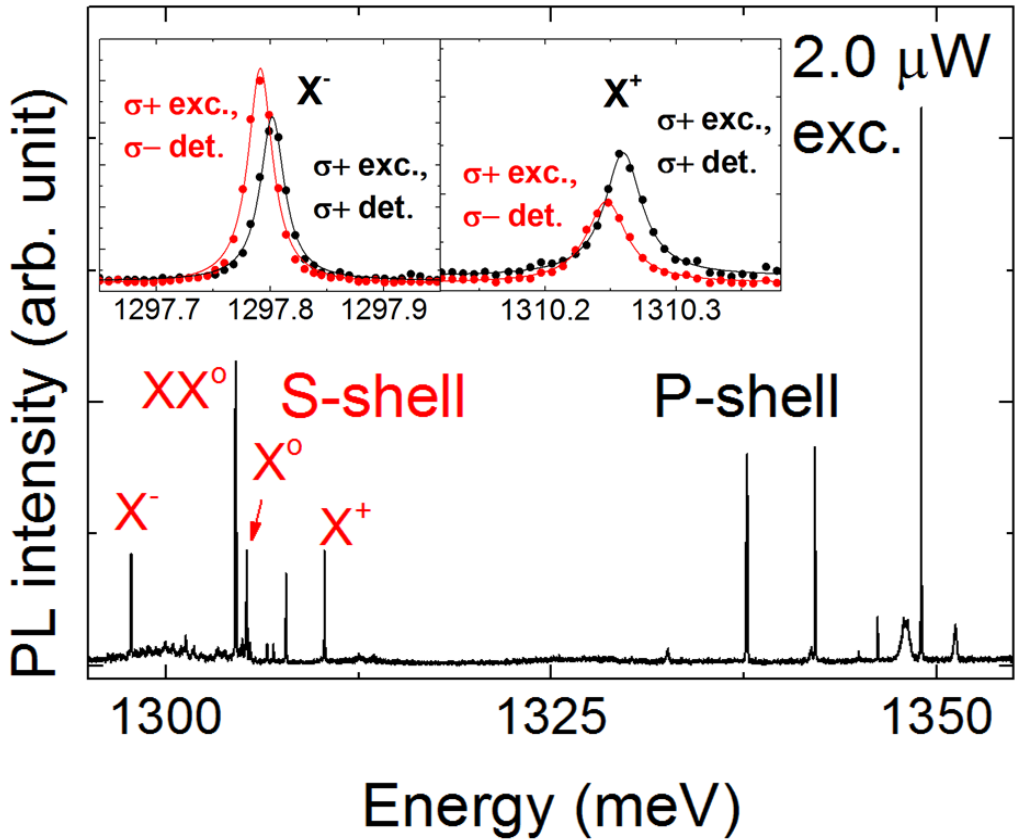

FIG. 1. (a) PL spectrum showing the s-shell and p-shell emission separated by about 50 meV. S-shell emission exhibits a number of peaks corresponding to $X^0$, $XX^0$, $X^+$ and $X^-$. Inset shows the cross- and co-polarized nature of $X^-$ and $X^+$ emission respectively. The separation between the peaks detected at orthogonal circular polarizations corresponds to the Overhauser shift.

Under polarization modulated excitation, the emission peak consists of contribution from both $\sigma^+$ and $\sigma^-$ excitation, each centered at a different energy separated by the OS. Since the OS is smaller than the linewidth of the emission peaks, these contributions superpose and thus give a single peak with a larger overall linewidth (Fig. 2(a)). As such we perform a two peak fit to the spectra and the separation of the two fitted peaks gives the OS. Analysis was performed on both $X^+$ and $X^-$, and we obtained similar and consistent results, showing that both charged excitons couple strongly to the nuclear field.

The results of $X^-$ are presented here. Shown in Fig. 2(a) is an example of a fitting for a spectrum taken at 1.5 µW excitation and 10 Hz modulation frequency, giving an OS of 10 µeV, comparable to previously reported values.[4] The key parameter in the two peak fitting is the width, which we obtain by measuring the linewidth of $X^-$ under DC excitation at the same power.[31]

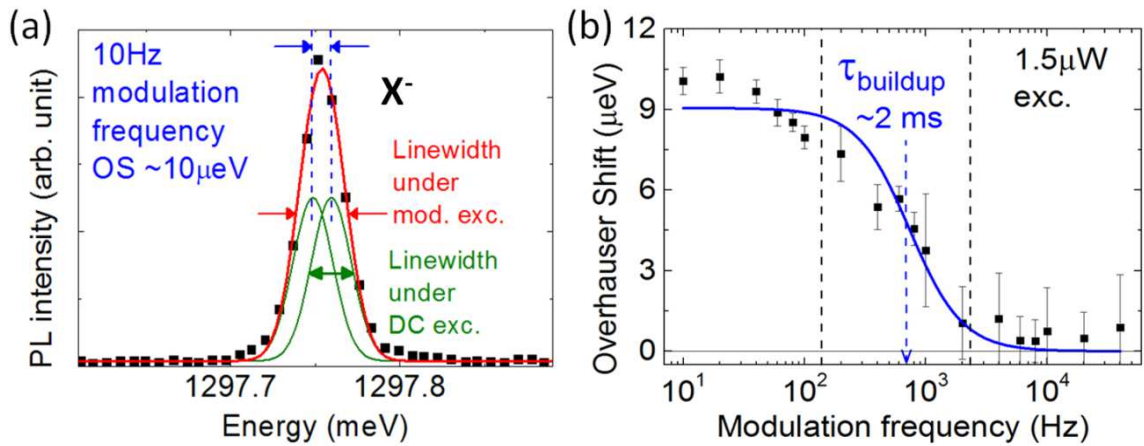

FIG. 2. *(a) Spectrum showing a two Gaussian fitting (green solid lines) to an $X^-$ peak where the separation of the fitted peaks give the OS. The red line gives the sum of the two fitted peaks. The respective linewidths under polarization modulation and DC excitation are as labelled. (b) The change in OS with modulation frequency allows us to*

*extract $T_{buildup}$ by fitting the data points with a Butterworth filter function. The dotted lines mark the three distinct regimes characteristic of such a measurement. The representative sample of data shown here indicates $T_{buildup}$ of about 2 ms at 1.5 µW excitation. The error bars represents the standard deviation of a number of data points taken at each frequency. The error in the value of OS could be induced by the instability of the position of the cryostat stage. The increasingly large error with modulation frequency is caused by the increasing uncertainty of the fitted peak position as the OS decreases.*

Figure 2(b) shows the behavior of the OS vs modulation frequency which can be considered to consist of three distinct regimes as marked by the dotted lines: at low modulation frequencies (<100 Hz), the OS is at its maximum (DC) value of about 10 µeV. As the frequency is low compared with $T_{buildup}$, the nuclei can follow the variation of the photo-modulated electron spin. Therefore, the nuclear spins are polarized to the fullest extent possible under the given experimental conditions. As the frequency increases, the measured OS reduces: each cycle of the modulation gets shorter and thus the nuclear spins get less polarized, resulting in weaker nuclear field and therefore smaller OS. At high frequencies (>1 kHz), the OS tends towards its minimum value indicating little or no nuclear spin polarization. At these frequencies, the electron spins switch so rapidly that the nuclear spin ensemble does not get polarized.

Based on the rate equation for the optical pumping of nuclear spin polarization,[2,32] we solve for the square wave polarization modulation excitation with frequency $\omega$, and obtained a solution in the form of the Butterworth filter function:

$\langle I_z \rangle = \alpha / \left( \omega^2 + \left( \frac{1}{T_{\text{buildup}}} \right)^2 \right)$, with $\langle I_z \rangle$ being the mean nuclear spin polarization and $\alpha$ is the amplitude fitting parameter to the spin polarization at no modulation (see section 5 and appendix for further details). By fitting this function to the data points, we could determine $T_{buildup}$. For the fitting process, we sometimes included a small constant offset in the fitting function in order to compensate for the fluctuation of the measured DC linewidths. The obtained $T_{buildup}$ is of the order of a few ms, which is consistent to previous reported values.[22,33]

$T_{buildup}$ takes the form $\frac{1}{T_{buildup}} = \frac{1}{T_{1e}} + \frac{1}{T_d}$, where it depends on the relative magnitude of two underlying timescales, namely the nuclear spin polarization time, $T_{1e}$ and nuclear spin decay time, $T_d$. These two timescales in turn depend on the experimental conditions, including but not limited to, the applied external magnetic field and the possible presence of a residual electron in QD.[10,22] It was found that a residual electron facilitates nuclear spin decay, leading to $T_{1e} > T_d$.[10] In our experiments, the sample is under CW excitation and thus we can assume that the QD could be occupied with a residual electron for a significant amount of time, leading to fast nuclear spin decay such that $T_{1e} > T_d$. As such, $T_{buildup}$ is more susceptible to changes in $T_d$, which supports the results of the power dependence of $T_{buildup}$ in the following section.

**IV. *P*-SHELL ASSISTED DNP**

The power dependence of the PL intensity of *s*- and *p*-shell emission (Fig. 3(a)) is measured by summing the integrated intensities of the peaks within 1297 – 1311 meV

(1337 – 1352 meV) of Fig. 1 for *s* (*p*)-shell. With increasing excitation power, the *s*-shell emission increases and then saturates, while the *p*-shell emission increases and eventually exceeds the *s*-shell emission intensity. In these high pumping-power conditions, the *s*-shell is closed and hence hinders the relaxation of *p*-shell carriers, which otherwise relax to the ground state within a picosecond timescale. The prolonged lifetime of *p*-shell carriers increases not only the radiative recombination but also their interaction with nuclear spins.

Figure 3(b) shows pump power dependences of the OS and $T_{buildup}$. The OS curve shows a continuous increase, even after the saturation of the *s*-shell emission, and reaches an OS of more than 13 µeV without any external magnetic field. $T_{buildup}$ show a gradual increase at low pump powers, which could arise from an increase of $T_{1e}$ due to suppressed electron-nuclear spin flip-flop processes by increased nuclear field[10,33] (which increase the energy mismatch between the electron spin states and hinders the flip-flop process). Then, $T_{buildup}$ shows an abrupt increase at excitation power above 1.5 µW, exactly when the *p*-shell begins to dominate.

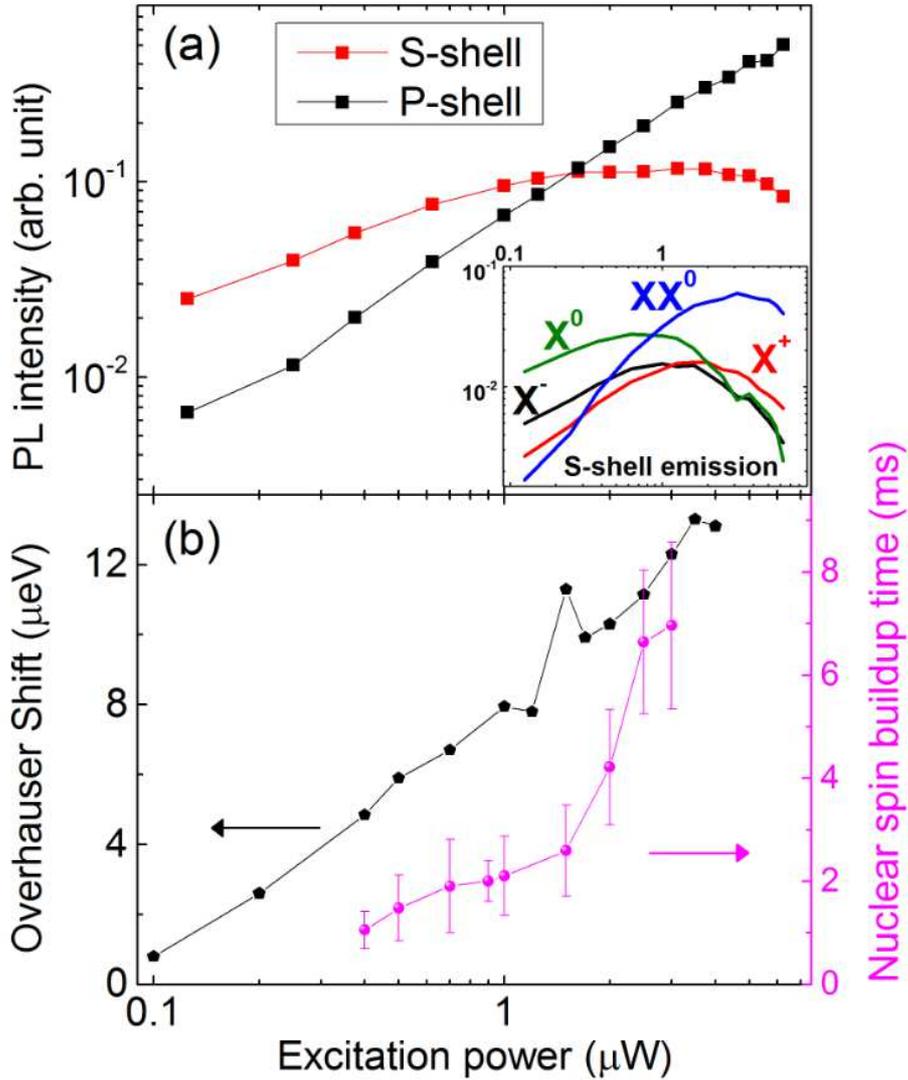

FIG. 3. (a) Plot showing the power dependence of s-shell and p-shell emission. The total PL intensity at each excitation power is obtained by summing the integrated intensities of peaks of s- and p-shell emission respectively. Inset shows the power dependence of four s-shell excitonic complexes. (b) The Overhauser shift (black) under DC excitation increases with excitation power while the nuclear spin buildup time (magenta) remains relatively short before an abrupt increase as the p-shell state emission begins to overtake that of s-shell emission at just under 2 µW. The error bars of the buildup time are the standard deviation of a number of measurements at each excitation power.

The observed continuous increase of the OS along with a sudden jump in $T_{buildup}$ at high pump powers can be attributed to slowed nuclear spin decay (increased $T_d$) and possibly hastened nuclear spin polarization (decreased $T_{1e}$). This is supported by numerical simulations (section 5) where we demonstrate that smaller $T_{1e}/T_d$ ratios result in larger OS: faster nuclear spin polarization and slower decay produce stronger nuclear fields.

The *p*-shell can support the suppression of the nuclear spin diffusion through the mechanism as explained below. A high spatial variation of *p*-shell electron wavefunction results in a strong inhomogeneity in the Knight field,[34] inducing energy mismatch between neighboring nuclei and resulting in the suppression of nuclear spin diffusion through dipole-dipole interaction.[1] The higher number of charged states of *p*-shell electrons and the greater degree of spatial variation of the *p*-shell could produce an even more strongly inhomogeneous Knight field. The inhomogeneous Knight field could lead to a quick rise in $T_d$ and thus $T_{buildup}$. To rule out DNP by delocalized carriers in the wetting layer, we note that these carriers do not suppress the nuclear spin diffusion as reported in reference [34] and thus do not support the observation of the sharp increase in $T_{buildup}$.

The *p*-shell could also contribute to nuclear spin polarization from two aspects. One is increased probability to have unpaired electrons, which could translate to a larger number of states that could induce DNP. Another is a larger spatial extension of the electron wavefunction than that of *s*-shell, which assists the nuclear spin polarization in the exterior of the *s*-shell wave function.[35] Overall, *p*-shell electrons could lead to larger nuclear spin polarization.

We consider that the increase of $T_d$ is predominantly responsible for the experimental observation. Although a decrease of $T_{1e}$ can explain the increase of OS (since $T_{1e}/T_d$ reduces), it cannot account for the increase of $T_{buildup}$ (given a fixed $T_d$). On the other hand, increase of $T_d$ can consistently explain both the observations ($T_{buildup}$ jump together with the increase of OS) and is considered to be the more likely scenario. Indeed, numerically estimated $T_{1e}$ is in excess of 30 ms, while $T_d$ is less than 10 ms (see also Fig. 4). As such any significant changes in OS and $T_{buildup}$ has to be due to changes in $T_d$.

We also rule out the possibility of a $T_d$ increase solely due to the closing of *s*-shell. At high pump powers with dominant *p*-shell emission, the *s*-shell orbital tends to be filled with paired electrons which do not disturb nuclear spins and hence result in less nuclear spin depolarization and thus longer $T_d$. However, even at high pump powers, there remains significant emission from neutral/charged excitons of the *s*-shell (Fig. 3(a) inset) which consist of unpaired electrons that interact with the nuclear spins. Furthermore, the residual electron after the recombination of X⁻ could facilitate depolarization as mentioned earlier. The combined effect of the polarization and depolarization by the *s*-shell excitonic complexes could at best give a small increase in $T_d$ as the *s*-shell closes. Moreover, the closed *s*-shell cannot efficiently polarize the nuclear field and hence cannot account for the continuous increase of the OS. Overall, there is less likelihood of $T_{buildup}$ increasing along with continuous increase of the OS due to the closing of *s*-shell. Therefore we propose that changes of the nuclear spin dynamics arise, not from the changes in the *s*-shell but from the interaction between *p*-shell electrons and nuclear spins in the QD.

## V. MODELLING AND SIMULATION

To support the above-mentioned interpretation of nuclear spin dynamics, we carried out simulations using a simple rate equation model based on earlier work,[10] originally proposed by Abragam,[32] given by

$$\frac{d\langle I_z^i\rangle}{dt} = -\frac{1}{T_{1e}}\left(\langle I_z^i\rangle - \frac{4}{3}I^i(I^i+1)\langle S_z\rangle\right) - \frac{1}{T_d}\langle I_z^i\rangle \tag{1}$$

where $\langle I_z^i\rangle$ is the mean nuclear spin polarization along the z axis, $I^i$ is the spin of the $i^{th}$ nucleus and $\langle S_z\rangle$ is the mean electronic spin along the z axis. The first term on the right-hand side of equation (1) describes the polarization of nuclear spins by electron spin, governed by timescale $T_{1e}$, and the second term describes the nuclear spin depolarization by timescale $T_d$. $T_{1e}$ takes the form of $T_{1e}^0\left\{1+\tau_{el}^2\left[\frac{g_{el}\mu_B}{\hbar}(B_{ext}+B_{nuc})\right]^2\right\}$ where $T_{1e}^0$ is the nuclear spin polarization time at zero total magnetic field, $\tau_{el}$ is the electron spin correlation time, $g_{el}$ is the electron g-factor, $\mu_B$ being the Bohr magneton, while $B_{ext}$ (= 0 in our case) and $B_{nuc}$ are the external and nuclear field respectively.

As circular polarization is transferred to the electron spin, to model the σ+/σ- polarization modulated square wave excitation, we introduced $\langle S_z\rangle = \langle S_z^0\rangle \frac{2}{i\pi}\sum_{n=1,odd}^{\infty}\frac{1}{n}\left(e^{ni\omega t}-e^{-ni\omega t}\right)$. The value of $\langle S_z^0\rangle$ is taken from the maximum degree of polarization up to ~0.2 (20%) measured under CW DC conditions.

As $T_{1e}$ is itself dependent on the nuclear field, this makes equation (1) nonlinear. However, assuming linear behaviour as has been done in previous reports,[5,10] we solve equation (1) to obtain a solution in the form of Butterworth filter function. We also solved equation (1) numerically, retaining its nonlinear features, in particular the dependence of $T_{1e}$ on $\langle I_z \rangle$ and we found that the two solutions are consistent (see appendix). For the comparison with the measured OS, we converted the simulated $\langle I_z \rangle$ to OS which are related by the relationship $OS = 2A\langle I_z \rangle$, where $A$ is the hyperfine constant, which is about 50 μeV for an InAs/GaAs QD of typical composition.[1,5]

Figure 4 shows a series of simulated OS as a function of $T_{1e}^0$ and $T_d$ under three different $\tau_{el}$ (all other parameters are fixed). It is apparent that the maximum OS essentially depends on the ratio $T_{1e}^0/T_d$. A small ratio reflects a high rate of polarization to decay and thus giving large OS while a large ratio gives the opposite. The resultant OS is also dependent on the electron correlation time, $\tau_{el}$ which describes the electronic spin state energy broadening. Increasing $\tau_{el}$ narrows the energy broadening which in turn decreases the probability of spin flips and therefore lowers the resulting nuclear spin polarization. However, regardless of the value of $\tau_{el}$, the regions which span the observed OS in the experiment indicates that $T_{1e}^0 > T_d$ as expected.

Matching the experimentally-observed OS to the simulation results, OS of 1 μeV to 13 μeV corresponds to $T_{1e}^0$ between 40 ms to 120 ms, while $T_d$ ranges from 2 ms to 6 ms, or possibly larger for both timescales. It is worth noting that unlike $T_{1e}^0$, $T_{1e}$ is magnetic field dependent such that with any magnetic field (in our case, nuclear field $B_{nuc}$), the value of $T_{1e}$ is always greater than $T_{1e}^0$. Given the relatively large $T_{1e}$, its reciprocal should remain relatively constant, therefore leaving $T_{buildup}$ to be easily affected by the increase in $T_d$, supporting experimental observation.

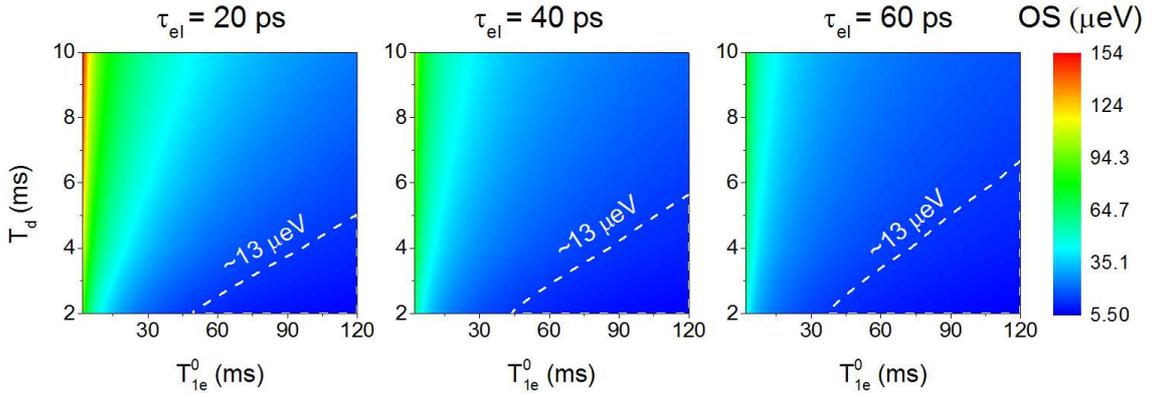

FIG. 4 *Color plot of the maximum OS obtained over a range of $T_{1e}^0$ and $T_d$ values for $\tau_{el} = 20, 40$ and $60\ ps$. Systems with short spin flip time and long nuclear spin decay time will give high OS corresponding to the top left corner of each plot. For higher values of $\tau_{el}$, the spin flip probability decreases and thus for the same values of $T_{1e}^0$ and $T_d$, the achievable $\langle I_z \rangle$ is less. The dashed line marks the approximate maximum OS observed in the experiments indicating that we are essentially operating in the regime where $T_{1e}^0 > T_d$.*

## VI. CONCLUSION

To conclude, we observed *p*-shell assisted DNP in QD at zero external magnetic field. We observed continued increase of the OS and a jump in $T_{buildup}$ as the *p*-shell emission becomes dominant. It was found that *p*-shell carriers are responsible for the increase in nuclear spin polarization after the saturation of the *s*-shell. The contribution of *p*-shell electrons to DNP is supported by measuring the power dependence of the nuclear spin buildup time. We consider that *p*-shell electrons slow down the nuclear spin diffusion by increasing the inhomogeneity of the Knight field. These in turn led to a continuous increase of OS after closing the *s*-shell together with the marked increase in the nuclear spin buildup time. The use of the *p*-shell also enables more nuclear spin polarization due to increased electron-nuclear spin interaction. Control over the population of the *p*-shell could allow us to break the current limit in nuclear spin polarization.

**ACKNOWLEDGEMENTS**

This work was supported by Project for Developing Innovation Systems of MEXT, Japan and JSPS KAKENHI Grant-in-Aid for Specially promoted Research (15H05700). We thank N. Kumagai and NEC Corporation for sample growth and preparation.

**APPENDIX: RESPONSE OF NUCLEAR SPIN UNDER CIRCULAR POLARIZATION MODULATED EXCITATION**

**Temporal and modulation frequency response of $I_z$**

Figure 5 shows the temporal response of the nuclear spin polarization, $I_z$ under square wave circular polarization modulated excitation. Despite the discrepancy between the magnitudes of the nuclear spin polarization for the solutions with and without linear approximation, both gave similar modulation of the nuclear spin polarization with the excitation. The overall behaviour where the nuclear spin polarization decreases with increasing modulation frequency can be clearly seen in the temporal behaviour.

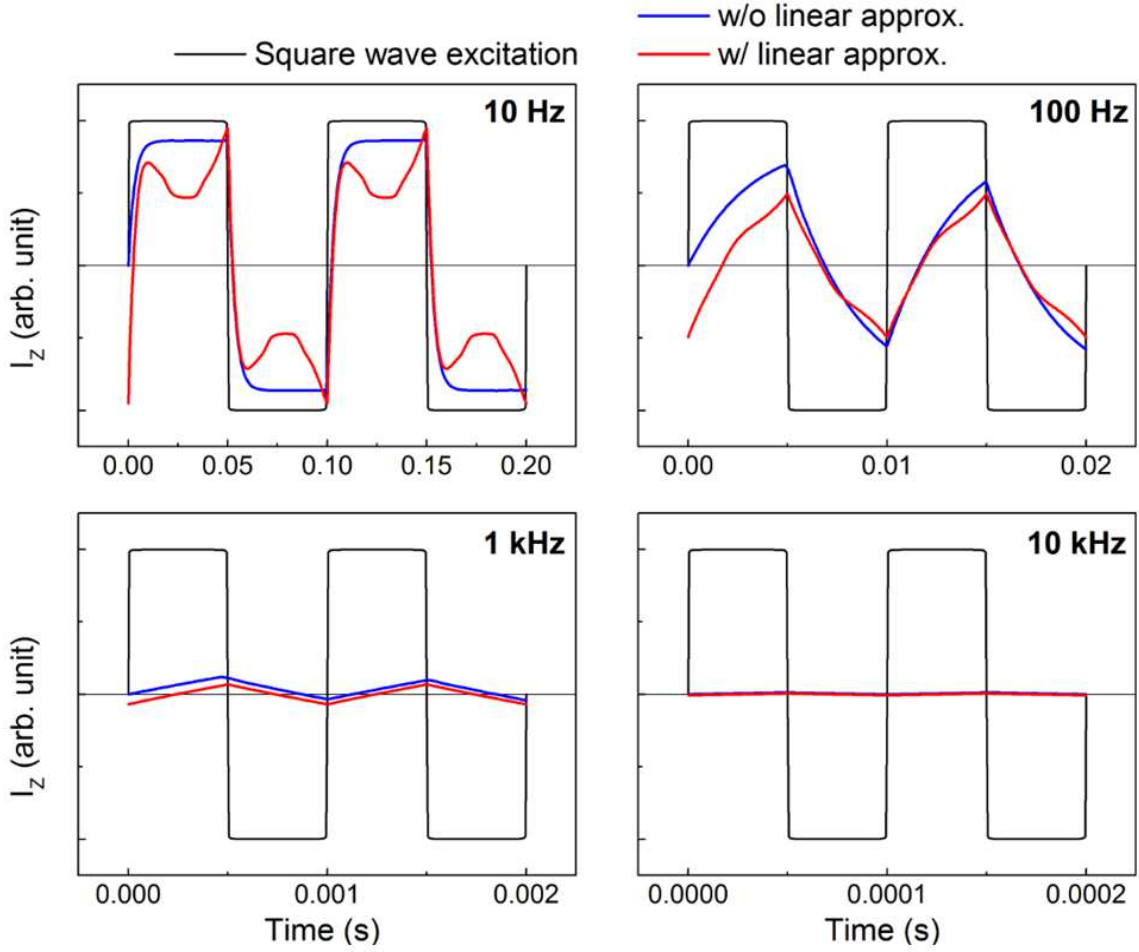

*FIG. 5 The temporal response of the nuclear spins is plotted against the square wave excitation (black lines) at modulation frequencies of 10 Hz, 100 Hz, 1 kHz, and 10 kHz for $T_{1e}^0 = 40\ ms, T_d = 4\ ms$ and $\tau_{el} = 60\ ps$. The blue (red) lines correspond to solutions of $I_z(t)$ without (with) linear approximation in equation 1 plotted on the same y-scale for all modulation frequencies. The two solutions are largely consistent with each other albeit the difference in the value of $I_z$. As the modulation frequency increases, the modulation amplitude of the nuclear spin polarization decreases as observed in the experiments.*

By summing the absolute values of $I_z$ over a period of time corresponding to the integration time (or alternatively summing values over a few periods to reduce computation time), we can obtain the time average value of $I_z$ i.e. $\langle I_z \rangle$ for each modulation frequency. The OS is proportional to $\langle I_z \rangle$. Simulation results of the change of OS with modulation frequency is consistent with that from experiments, allowing us to conclude that the linearization assumption is valid, so are the analytical solutions of $I_z(t)$ and $T_{buildup}$.

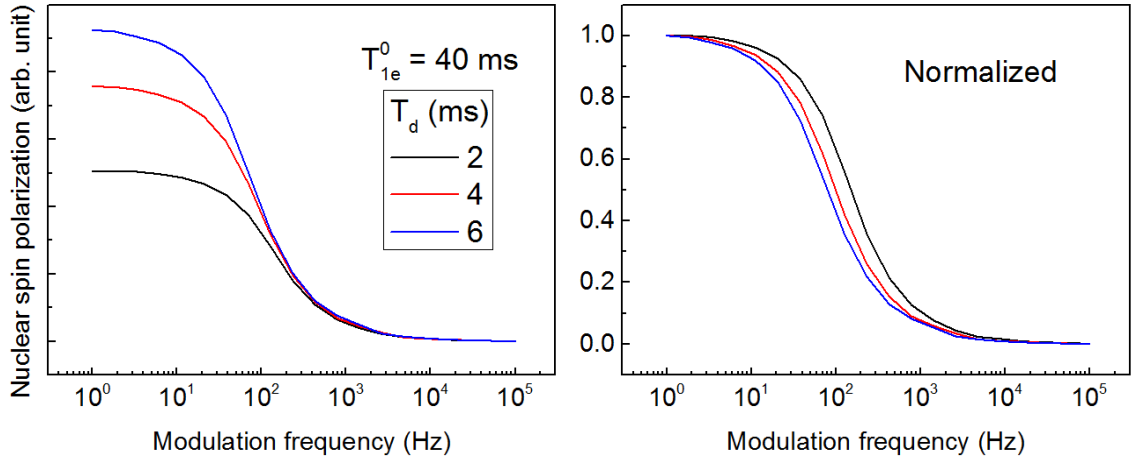

FIG. 6 *The plots show the change of the nuclear spin polarization with modulation for $T_d = 2, 4$ and $6$ ms without linear approximation. Other parameters are fixed at $T_{1e}^0 = 40$ ms and $\tau_{el} = 60$ ps. As $T_d$ increases (ratio $T_{1e}^0/T_d$ decreases) nuclear spin polarization starts to decrease at lower modulation frequency, meaning longer $T_{buildup}$.*

For a fixed value of $T_{1e}^0 = 20$ ms, Fig. 6 shows how the nuclear spin polarization response to modulation frequency changes for different values of $T_d$. For longer $T_d$, there is less nuclear spin diffusion per unit time and thus the maximum achievable nuclear spin polarization at low modulation frequency is higher. The normalized plots show how the nuclear spin polarization starts to decrease at lower frequency for longer buildup times and vice versa.